\documentstyle[aps,prc,twocolumn]{revtex}    

\tighten

\def\sfrac#1#2{{\textstyle \frac{#1}{#2}} }
\begin{document}

\preprint{ }
\draft

\wideabs{      

\title{Propagation of a massive spin-3/2 particle}

\author{Helmut Haberzettl}

\address{Center for Nuclear Studies, Department of Physics,\\ 
The George Washington University, Washington, D.C. 20052}

\date{16 December 1998}
\maketitle                                                                             %

\begin{abstract}
The propagation of an off-shell spin-3/2 baryon is investigated in a consistent framework 
using projections corresponding to one irreducible spin-3/2 and two irreducible spin-1/2 
representations arising from the usual spinor-vector representation of the spin-3/2 fields.
Starting from the most general Lagrangian invariant under point transformations containing 
first-order derivatives only and imposing the constraint $\gamma_\mu \psi^\mu=0$, we 
eliminate one of the spin-1/2 contributions. The resulting propagator provides a 
clean separation of spin-3/2 proper and the one remaining spin-1/2 contribution. In 
addition to a conventional pole term describing pure spin-3/2, which is identical to the
propagator proposed by Williams, this procedure yields a second pole term describing 
spin-1/2 only, with a pole at twice the mass parameter of the Lagrangian. Its effect on 
physical observables may manifest itself in a manner which is indistinguishable from a 
particle in its own right. The second pole term cancels the $1/p^2$ singularity of the 
Williams propagator and the propagator derived here is well behaved for vanishing $p^2$. 
\end{abstract}
\pacs{PACS numbers: 11.10.Ef, 14.20.Gk, 13.75.Gx \hfill {\bf nucl-th/9812043}  [4]} 
}    

\section{Introduction}
The description of higher spin fields is complicated by the well-known fact that a field with a given spin $s$
($\ge 1$), in addition to the spin $s$ of interest, will also contain lower spin components $(s-1)$, $(s-2)$, etc.,
and that the description of these components is not unique. 
With electron accelerator facilities like CEBAF at the
Jefferson Laboratory coming on-line, and the pressing need to accurately account for the structure of hadrons in the
resonance regions in order to be able to interpret the data, it is becoming increasingly important to find some consensus 
among theorists as to how to
correctly describe resonant states with higher spins. One of the basic prerequisites in this respect is a
description of the spin-3/2 resonances, like the $\Delta(1232)$. 
 This problem has
generated a lot of discussions in the literature (see Refs.\ \cite{RS,spin32,peccei,nieuw,williams,benmer,pascalutsa}
and references therein), and it is still not settled.

Recently, a very interesting approach to spin-3/2 fields was put forward by Pascalutsa \cite{pascalutsa}, based on a Dirac
constraint analysis of the Rarita--Schwinger Lagrangian within the path-integral approach,
aiming at restricting the description to spin-3/2 components proper only. Pascalutsa's results show that
for the Rarita--Schwinger Lagrangian, the spin-1/2 contributions cannot be eliminated
at the level of the free field. He finds that restricting the description of 
the propagating off-shell baryon to spin-3/2 contributions only requires
imposing the constraints at the level of the interaction
Lagrangian and that this necessarily involves introducing higher than first-order derivative couplings.

The present treatment of spin-3/2 fields is based on the most general Lagrangian
invariant under point transformations employing first-order derivatives only.
We do not aim here to eliminate the spin-1/2 contributions altogether.
Rather we want to identify their origins and see how much of them we may safely discard at the outset
without leading to unwanted, unphysical singularities.

In Sec.\ \ref{sec:spinvec}, we investigate the various spin contributions arising from the
usual spinor-vector representation of the spin-3/2 fields.
We show that the corresponding irreducible representations can be 
classified---appropriately for the present purpose---by whether 
$p_\mu \psi^\mu=0$ or $\gamma_\mu \psi^\mu=0$ is chosen as the primary constraint to be satisfied
by the spinor-vector field $\psi^\mu$.  It is demonstrated, in Sec.\ \ref{sec:lagrange}, that the 
usual Rarita--Schwinger construction of the propagator incorporates all spin-1/2 contributions. 
Going through an undefined $\frac{0}{0}$ situation due to `forbidden' parameters of the point transformation,
we show that one can easily eliminate the spin-1/2 contributions which do not satisfy the Rarita--Schwinger constraint 
$\gamma_\mu \psi^\mu=0$, without incurring any singularities. 

The resulting propagator exhibits a conventional pole term, identical to the Williams propagator \cite{williams},
and an additional pole at twice the mass $M$ defined by the first pole. This second term is found to cancel
the $1/p^2$ singularity of the numerator tensor of the Williams propagator and the resulting
combined expression admits a well-behaved limit for vanishing $p^2$. 
The second pole term cleanly isolates the spin-1/2 contributions from spin-3/2 proper. In the concluding 
Sec.\ \ref{sec:sum}, we discuss the ramifications of this finding for physical observables.

\section{Spinor-Vector Representations}\label{sec:spinvec}
First, 
let us recapitulate some well-known facts regarding the spin-3/2 spinor representation.
Denoting the basic spin-1/2 spinors of the $SU(2) \otimes SU(2)$ spinor representation
in the usual way (see, e.g.,  Ref.\ \cite{weinberg}) by
%
\begin{equation}
(\sfrac{1}{2},0) \quad\quad\text{and}\quad\quad (0,\sfrac{1}{2})  \;, 
\end{equation}
higher spin fields are constructed by taking tensor products between the spinors. A
spin-1 vector spinor is then given by
%
\begin{equation}
\text{Vector: }\quad   (\sfrac{1}{2},0) \otimes (0,\sfrac{1}{2}) = (\sfrac{1}{2},\sfrac{1}{2})\;, 
\end{equation}
and a spin-3/2 spinor is usually, according to the Rarita--Schwinger construction \cite{RS}, introduced as
the product of a vector and a spin-1/2 spinor, i.e.,
\begin{equation}\label{spin32_decompose}
\text{Spin-3/2: }\quad   (\sfrac{1}{2},\sfrac{1}{2}) \otimes (\sfrac{1}{2},0) = (1,\sfrac{1}{2}) \oplus (0,\sfrac{1}{2})\;, 
\end{equation}
which is usually written as a four-spinor $\psi^\mu$ with a vector index $\mu$. 
[We conciously avoid writing Eq.\ (\ref{spin32_decompose}) in the more correct way as
\[(\sfrac{1}{2},\sfrac{1}{2}) \otimes \left[ (\sfrac{1}{2},0) \oplus (0\sfrac{1}{2})\right] 
= (1,\sfrac{1}{2}) \oplus (0,\sfrac{1}{2})\oplus (\sfrac{1}{2},1) \oplus (\sfrac{1}{2},0)\]
since this properly symmetrized form only complicates the notation, without adding anything 
substantial to the present discussion.]

As is seen from Eq.\ (\ref{spin32_decompose}), this field has one spin-3/2 component proper contained in $(1,\sfrac{1}{2})$.
In addition, there are
two spin-1/2 contributions, one from the Dirac spinor $(0,\sfrac{1}{2})$ and
one from combining the spins $1$ and $\frac{1}{2}$ in $(1,\sfrac{1}{2})$ to a total spin $\frac{1}{2}$.
As a consequence, the spinor-vector representation gives rise to one irreducible spin-3/2
and two irreducible spin-1/2 representations.

In the following we will classify two different ways of introducing explicit representations by the 
way one describes the Dirac $(0,\sfrac{1}{2})$ field contribution in the direct sum, Eq.\ (\ref{spin32_decompose}).

\subsection{Using \boldmath{$p_\mu \psi^\mu=0$} as the primary constraint}

In view of the fact that it has the same transformation properties as a Dirac field, we may identify
$p^\mu \psi_\mu$ with $(0,\sfrac{1}{2})$, i.e.,
\begin{equation}\label{irred_def_p}
(0,\sfrac{1}{2}) = p^\mu \psi_\mu \;, 
\end{equation}
where $p^\mu$ is the four-momentum of the particle described by the field $\psi^\mu$. The complementary
component is then given by
\begin{equation}\label{dspace_p}
(1,\sfrac{1}{2}) = \left(g^{\mu\nu}-\frac{p^\mu p^\nu}{p^2} \right)  \psi_\nu \;, 
\end{equation}
which has a zero contraction with $p_\mu$. 
It is this latter part of the spinor-vector space we are interested in since it contains spin-3/2.
Equation (\ref{dspace_p}) obviously amounts to employing $p_\mu \psi^\mu=0$ as
a primary constraint for the field which must be satisfied regardless of whether we are on-shell or not.

To isolate the spin-1/2 contribution still contained within $(1,\sfrac{1}{2})$, one must construct
a projector whose contraction with $p^\mu$ is zero, but whose contraction with $\gamma^\mu$ must not be zero.
One easily finds
%
\begin{equation}
g^{\mu\nu}-\frac{p^\mu p^\nu}{p^2}= {\cal D}^{\mu\nu}+({\cal P}_{11})^{\mu\nu}  \;, 
\end{equation}
where
\begin{equation}\label{proj32}
{\cal D}^{\mu\nu}    =  g^{\mu\nu} -\frac{p^\mu p^\nu}{p^2} 
             -\frac{(p^\mu-\gamma^\mu {p\!\!\!/})(p^\nu-{p\!\!\!/}\gamma^\nu )}{3p^2}
\end{equation}
is the projection operator associated with spin-3/2 proper and
%
\begin{equation}
({\cal P}_{11})^{\mu\nu} = \frac{(p^\mu-\gamma^\mu {p\!\!\!/})(p^\nu-{p\!\!\!/}\gamma^\nu )}{3p^2} 
\end{equation}
is the desired projector onto the spin-1/2 component of $(1,\sfrac{1}{2})$ in the irreducible representation
defined by Eq.\ (\ref{irred_def_p}), with
\begin{equation}
({\cal P}_{22})^{\mu\nu} = \frac{p^\mu p^\nu}{p^2}  \; 
\end{equation}
being the projector associated with $(0,\frac{1}{2})$.

The operators \cite{nieuw}
\begin{mathletters}
\begin{eqnarray}
({\cal P}_{12})^{\mu\nu} &=& -\frac{(p^\mu-\gamma^\mu{p\!\!\!/} )p^\nu}{\sqrt{3}p^2}  \;, \\
({\cal P}_{21})^{\mu\nu} &=& -\frac{p^\mu(p^\nu-{p\!\!\!/}\gamma^\nu )}{\sqrt{3}p^2}   \; 
\end{eqnarray}
\end{mathletters}
describe the transitions from one irreducible spin-1/2 representation to the other and allow one to
mix the two representations. Evidently, ${\cal D}$, ${\cal P}_{11}$, and ${\cal P}_{22}$ are
mutually orthogonal and
%
\begin{equation}
g^{\mu\nu} = {\cal D}^{\mu\nu}+({\cal P}_{11})^{\mu\nu}+({\cal P}_{22})^{\mu\nu}  \; 
\end{equation}
provides an expansion of the identity.

These projection operators are well-known and form the basis of the investigations of spin-3/2 fields
in Refs.\ \cite{spin32,peccei,nieuw,williams,benmer}. 
The derivation given here  makes it obvious that this particular set of
irreducible representations suffers from an unphysical singularity at $p^2=0$ which---most importantly---affects
\begin{equation}
 g^{\mu\nu}-\frac{p^\mu p^\nu}{p^2}  \;,
\end{equation}
which serves as the metric tensor for the $(1,\frac{1}{2})$ spinor-vector subspace [cf.\ Eq.\ (\ref{dspace_p})].

\subsection{Using \boldmath{$\gamma_\mu \psi^\mu=0$} as the primary constraint}

As an obvious alternative to Eq.\ (\ref{irred_def_p}), we may make the identification
\begin{equation}\label{irred_def_g}
(0,\sfrac{1}{2}) = \gamma^\mu \psi_\mu \; 
\end{equation}
instead, since $\gamma^\mu \psi_\mu$ also transforms the same way as a Dirac spinor.
Its complementary component,
\begin{equation}\label{dspace}
(1,\sfrac{1}{2}) = \left(g^{\mu\nu}-\sfrac{1}{4} \gamma^\mu \gamma^\nu\right)  \psi_\nu \;, 
\end{equation}
then has zero contraction with a $\gamma_\mu$ matrix, i.e., we now have
implemented $\gamma_\mu \psi^\mu=0$ as the primary constraint for the $(1,\sfrac{1}{2})$ field contributions.
This representation of $(1,\sfrac{1}{2})$ is referred to as the Rarita--Schwinger (RS) field \cite{RS,weinberg}.
Evidently, the metric tensor of the associated subspace,
\begin{equation}\label{dmetric}
g^{\mu\nu}-\sfrac{1}{4} \gamma^\mu \gamma^\nu \;,
\end{equation}
is nonsingular (and momentum-independent), which immediately removes one of the major objections against the
previous irreducible representations.

Introducing the projection operators,
\begin{mathletters}
\begin{eqnarray}
{\sf D}^{\mu\nu} &=& g^{\mu\nu}-\sfrac{1}{4} \gamma^\mu \gamma^\nu  \label{projd}\;,\\
{\sf P}^{\mu\nu}&=& \sfrac{1}{4} \gamma^\mu \gamma^\nu   \label{projp}\;,
\end{eqnarray}
\end{mathletters}
the spin-3/2 field can then be decomposed according to
\begin{equation}\label{pdpsi}
\psi^\mu =\psi^\mu_{\sf D}+\psi^\mu_{\sf P}={\sf D}^{\mu\nu} \psi_\nu+{\sf P}^{\mu\nu} \psi_\nu   \;, 
\end{equation}
where
\begin{equation}
\psi^\mu_{\sf P}={\sf P}^{\mu\nu} \psi_\nu   \; 
\end{equation}
contains no spin-3/2 components. To identify the spin-1/2 components within the RS field,
\begin{equation}
\psi^\mu_{\sf D}={\sf D}^{\mu\nu} \psi_\nu\; ,
\end{equation}
we write
%
\begin{equation}
{\sf D}^{\mu\nu} = {\cal D}^{\mu\nu}+\overline{{\sf P}}^{\mu\nu}\;,\\
\end{equation}
where 
\begin{equation}
\overline{{\sf P}}^{\mu\nu}=\frac{4(p^\mu-\frac{1}{4}\gamma^\mu {p\!\!\!/})(p^\nu-\frac{1}{4}{p\!\!\!/}\gamma^\nu )}{3p^2}\;,\\
\end{equation}
projects onto the subspace associated with that spin-1/2 field. ${\cal D}^{\mu\nu}$ is the
spin-3/2 projector of Eq.\ (\ref{proj32}) which we can also write as
\begin{equation}\label{proj32_alt}
{\cal D}^{\mu\nu}    =  g^{\mu\nu} -\sfrac{1}{4}\gamma^\mu \gamma^\nu 
              - \frac{4(p^\mu-\frac{1}{4}\gamma^\mu {p\!\!\!/})(p^\nu-\frac{1}{4}{p\!\!\!/}\gamma^\nu )}{3p^2} \;,            
\end{equation}
in order to make the connection to the present treatment more obvious.

${\cal D}$, ${\sf P}$, and $\overline{\sf P}$ are mutually orthogonal and therefore ${\sf P}$ and $\overline{\sf P}$ 
correspond to alternative
irreducible representations of the two spin-1/2 fields arising within the spinor-vector representation, i.e.,
%
\begin{equation}
{\sf P} + \overline{\sf P} = {\cal P}_{11} + {\cal P}_{22}\;. 
\end{equation}
In this context, we mention that 
\begin{eqnarray}
\overline{\sf P}&=& {\sf D}\left( {\cal P}_{11}+{\cal P}_{22} \right){\sf D} \nonumber\\ 
                &=& \sfrac{1}{4}{\cal P}_{11}+\sfrac{3}{4}{\cal P}_{22}
                     -\sfrac{\sqrt{3}}{4}{\cal P}_{12}-\sfrac{\sqrt{3}}{4}{\cal P}_{21}
\end{eqnarray}
provides an explicit description of how the two irreducible representations given in the previous subsection are mixed
in the present representations. 

In view of the nonsingularity of the metric associated with $(1,\frac{1}{2})$ [cf.\ Eq.\ (\ref{dmetric})],
we will use the spin-1/2 representations introduced here in the following.

\section{Lagrangians and Propagators}\label{sec:lagrange}                   
We recall that the free Lagrangian for the spin-3/2 case takes the form
%
\begin{equation}\label{Lagrange}
{\cal L}= \overline{\psi}_\mu \Lambda^{\mu\nu} \psi_\nu^{\,}  \;,
\end{equation}
which, using Eq.\ (\ref{pdpsi}), we can rewrite in the matrix form
\begin{equation} \label{matrixlagrange}
{\cal L} =  \left( \begin{array}{cc} 
   \overline{\psi}{\sf D} \;&\; \overline{\psi} {\sf P}
   \end{array} \right)_\mu 
    \left( \begin{array}{cc} 
       {\sf D}\Lambda {\sf D}   &  {\sf D}\Lambda {\sf P}   \\[2mm]
       {\sf P}\Lambda {\sf D}   &  {\sf P}\Lambda {\sf P}          
    \end{array} \right)^{\mu\nu} 
    \left( \begin{array}{c} 
       {\sf D}\psi  \\[2mm]
       {\sf P}\psi
    \end{array} \right)_\nu^{\,}  \;. 
\end{equation}
This explicitly shows that this Lagrangian will have spin-1/2 pieces 
originating from $(0,\frac{1}{2})$ unless
${\sf D}\Lambda {\sf P}={\sf P}\Lambda {\sf D}={\sf P}\Lambda {\sf P}=0$.

The most general form for $\Lambda^{\mu\nu}$ containing up to first-order derivatives only
is given by \cite{spin32,benmer}
%
\begin{eqnarray}\label{genlagrange}
\Lambda^{\mu\nu} &=& ({p\!\!\!/}-M)g^{\mu\nu} +A(\gamma^\mu p^\nu+p^\mu\gamma^\nu)  \nonumber\\
& &{ } +\sfrac{1}{2}(3A^2+2A+1)\gamma^\mu{p\!\!\!/}\gamma^\nu    \nonumber\\
& &{ } +M(3A^2+3A+1)\gamma^\mu\gamma^\nu  \label{ltensor} \;,
\end{eqnarray}
where $p^\mu = i \partial^\mu$ and $M$ is the mass of the spin-3/2 baryon and $A$ is an arbitrary parameter.

This particular form of the Lagrangian is obtained by demanding that it be invariant
under point transformations \cite{spin32},
\begin{mathletters}\label{point}
\begin{eqnarray}
\psi^\mu \rightarrow \psi'^\mu &=& \left(g^{\mu\nu}+a\gamma^\mu\gamma^\nu \right)\psi_\nu  \;, \label{pointpsi}\\
A        \rightarrow A'        &=& \frac{A-2a}{1+4a}  \;, \label{pointa}
\end{eqnarray}
\end{mathletters}
where $a$ is an arbitrary parameter, except that $a=-\frac{1}{4}$ is excluded since it would render the transformation
(\ref{pointa}) singular. For a similar reason, $A=-\frac{1}{2}$ is not allowed since the resulting
propagator would become infinite  [cf.\ Eq.\ (\ref{GRSA})].

The simplest interaction Lagrangian compatible with chiral symmetry and consistent with the present
approach contains a first-order derivative coupling which, considering the example of the
$\Delta N \pi$ vertex, leads to the coupling operator
\cite{spin32,benmer} 
%
\begin{equation}\label{oldvertex}
\Gamma^\mu(q;A,z) = \frac{f}{m_\pi} \theta^{\mu\nu}(A;z) q_\nu  \;, 
\end{equation}
where  $q$ and $m_\pi$ are the pion's momentum and mass, respectively, $f$ is the coupling strength
(and isospin is ignored for simplicity). The tensor
%
\begin{equation}
\theta^{\mu\nu}(A;z)= g^{\mu\nu} +\left[\sfrac{1}{2}(1+4z)A+z\right] \gamma^\mu \gamma^\nu   \; 
\end{equation}
contains a parameter $z$ measuring the `off-shellness' of the vertex.
Theoretical attempts to determine $z$ have been inconclusive so far and it is
widely thought that $z$ must be determined from experiment (see \cite{benmer},
and references therein).

\subsection{Rarita--Schwinger case}\label{sec:RS}

The well known Rarita--Schwinger (RS) Lagrangian is obtained with the choice $A=-1$, i.e.,
%
\begin{eqnarray}\label{lagrangers}
\Lambda^{\mu\nu}\rightarrow \Lambda^{\mu\nu}_{\text{RS}}&=& g^{\mu\nu}({p\!\!\!/}-M) 
             - ( \gamma^\mu p^\nu   +  p^\mu\gamma^\nu ) \nonumber\\
& &{ }       +\gamma^\mu {p\!\!\!/} \gamma^\nu +M \gamma^\mu\gamma^\nu  \nonumber\\
&=&{ }       \sfrac{i}{2} \left\{ \sigma^{\mu\nu} , ({p\!\!\!/}-M) \right\}  \;,
\end{eqnarray}
which, when solving \cite{spin32,benmer}
\begin{equation}\label{solveproprs}
\Lambda_{\mu\rho} G^{\rho\nu}_{\text{RS}} = g_\mu^{\:\:\nu}   \; 
\end{equation}
in momentum space, leads to the propagator 
\begin{equation}\label{proprs}
G^{\mu\nu}_{\text{RS}}  = \frac{({p\!\!\!/}+M)\Delta^{\mu\nu}_{\text{RS}}}{p^2-M^2}     \;, 
\end{equation}
with
\begin{equation} \label{deltars}
\Delta^{\mu\nu}_{\rm RS} = g^{\mu\nu} -\sfrac{1}{3} \gamma^\mu \gamma^\nu 
             -\frac{2p^\mu p^\nu}{3M^2}  -\frac{\gamma^\mu p^\nu -\gamma^\nu p^\mu}{3M} \;. 
\end{equation}
The latter is the usual Rarita--Schwinger propagator tensor. 

We would like to emphasize here that, as we have seen already from Eq.\ 
(\ref{matrixlagrange}), the RS tensor in general will admit {\it both} the
spin-1/2 contributions from $(1,\frac{1}{2})$ and $(0,\frac{1}{2})$ due
to the fact that Eq.\ (\ref{solveproprs}) seeks a
propagator that is obtained by inverting $\Lambda_{\mu\rho}$ on the entire space.

The RS vertex is given by
\begin{eqnarray}\label{RSvertex}
\Gamma^\mu_{\text{RS}}(q;z) &=& \frac{f}{m_\pi} \theta^{\mu\nu}(-1;z) q_\nu \nonumber\\
                  &=& \frac{f}{m_\pi} \left[g^{\mu\nu} - \left(\sfrac{1}{2}+z \right)\gamma^\mu \gamma^\nu\right] q_\nu \;, 
\end{eqnarray}
which still contains the undetermined parameter $z$.

Finally, we mention that the reason for choosing $A=-1$ is that the general solution of Eq.\ (\ref{solveproprs})
without prior choice of $A$ generates an additional $A$-dependent contact term of the
form \cite{spin32,benmer}
%
\begin{eqnarray}\label{GRSA}
G_A^{\mu\nu} &=& -\frac{1}{3M^2}\frac{A+1}{(2A+1)^2}  \nonumber\\
& & \times   \Big[(2A+1)(\gamma^\mu p^\nu + p^\mu\gamma^\nu) \nonumber\\
& &\quad\quad{ }- \sfrac{A+1}{2}\gamma^\mu ({p\!\!\!/}+2M)\gamma^\nu +M\gamma^\mu\gamma^\nu \Big]  \;, 
\end{eqnarray}
which must be added to $G^{\mu\nu}_{\text{RS}}$, and the RS choice $A=-1$ makes this term vanish. 
For $A=-\frac{1}{2}$, $G_A^{\mu\nu}$ becomes infinite.

\subsection{`Singular' case}

As indicated, the treatment of the previous subsection takes into account both spin-1/2 contributions.
However, since our primary interest are the proper spin-3/2 contributions only, and
unless one is forced to do so by the mathematics of the problem, 
there is really no need to take into account these extraneous spin-1/2 pieces
which are only an artifact of the spinor-vector construction.
We, therefore, would like to be able to restrict the description to $(1,\frac{1}{2})$ instead.
Eliminating the $(0,\frac{1}{2})$ pieces will have the added advantage that the primary constraint 
$\gamma_\mu \psi^\mu=0$ will be valid at all times.

To achieve this goal, the key here are the `forbidden' parameter values $A=-\frac{1}{2}$ and $a=-\frac{1}{4}$.
Note that for these values, firstly, the point transformation (\ref{pointpsi}) corresponds to
a projection onto the $(1,\frac{1}{2})$ RS spinor-vector space according to Eq.\ (\ref{projd})  and, secondly,
the mapping of $A$, Eq.\ (\ref{pointa}), leads to an undefined $\frac{0}{0}$ situation.

To investigate the latter situation in more detail, consider a value of $a$ related to $A$ by
%
\begin{equation}\label{arelation}
A = -\frac{1}{2}+(1+4a)^{1+x}  \;, 
\end{equation}
which means that Eq.\ (\ref{pointa}) becomes
\begin{equation}
A        \rightarrow A'        = -\frac{1}{2}+(1+4a)^x  \;, \label{pointaa}
\end{equation}
where any value of $x > 0$ is acceptable as long as (\ref{arelation}) has real solutions. 
This shows that for $a=-\frac{1}{4}$ and $A=-\frac{1}{2}$, the $\frac{0}{0}$ situation
is resolved and conforms to the transformations
\begin{mathletters}\label{point12}
\begin{eqnarray}
\psi^\mu \rightarrow \psi'^\mu &=& \left(g^{\mu\nu}-\sfrac{1}{4}\gamma^\mu\gamma^\nu \right)\psi_\nu \nonumber\\
&=&  {\sf D}^{\mu\nu} \psi_\nu  =  \psi^\mu_{\sf D} \;, \label{pointspsi}\\
A        \rightarrow A'        &=& -\frac{1}{2}  \;. \label{pointsa}
\end{eqnarray}
\end{mathletters}
This clearly means that going through the `singular' set of values, one restricts the description
to the  $(1,\frac{1}{2})$ RS fields, as desired.
Any subsequent point transformation will not lead out of this space since $(g_{\mu\rho}+a\gamma_\mu\gamma_\rho){\sf D}^{\rho\nu}
={\sf D}_\mu^{\;\;\nu}$ for any $a$. This, therefore, corresponds to a trivial realization of the invariance under the 
transformations (\ref{point}) in the sense that neither $\psi^\mu$ nor $A$ changes.

The corresponding $\Lambda^{\mu\nu}$ tensor is
\begin{eqnarray} \label{lagrange32}
\Lambda^{\mu\nu}&=& g^{\mu\nu}({p\!\!\!/}-M) 
             - \sfrac{1}{2}( \gamma^\mu p^\nu   +  p^\mu\gamma^\nu ) \nonumber\\
& &{ }       +\sfrac{3}{8}\gamma^\mu {p\!\!\!/} \gamma^\nu +\sfrac{1}{4} M \gamma^\mu\gamma^\nu  \;,
\end{eqnarray}
which follows from Eq.\ (\ref{ltensor}) with $A=-\frac{1}{2}$. With this tensor structure for the Lagrangian, 
it is then a trivial exercise to show that the $(0,\frac{1}{2})$ elements of (\ref{matrixlagrange}) indeed vanish,
as desired, i.e.,
\begin{equation}\label{pzero}
(\Lambda {\sf P})^{\mu\nu}=({\sf P}\Lambda)^{\mu\nu}=0   \;, 
\end{equation}
and the resulting Lagrangian,
\begin{equation}\label{hhlagrange}
{\cal L} = (\overline{\psi}{\sf D})_\mu({\sf D}\Lambda {\sf D})^{\mu\nu}({\sf D}\psi)_\nu   \;, 
\end{equation}
obviously only describes RS field contributions.

To determine the propagator for this case, we need to solve
%
\begin{equation}\label{solveprop}
({\sf D}\Lambda {\sf D})_{\mu\rho} G^{\rho\nu} = {\sf D}_\mu^{\;\;\nu}   
\end{equation}
in momentum space, which replaces Eq.\ (\ref{solveproprs}). It is crucial here that
we seek only an inverse of ${\sf D}\Lambda {\sf D}$ on the $(1,\frac{1}{2})$ spinor-vector space
since this is the only part of the space where $\Lambda$ has non-zero elements.

The resulting propagator is
%
\begin{equation}\label{myprop}
G^{\mu\nu}  = \frac{({p\!\!\!/}+M) {\cal D}^{\mu\nu} }{p^2-M^2}+\frac{{\cal N}^{\mu\nu} }{p^2-(2M)^2}  \;,
\end{equation}
where the first term contains the spin-3/2 projection operator, Eq.\ (\ref{proj32_alt}),
satisfying ${\sf P}{\cal D}={\cal D}{\sf P}=0$ and ${\sf D}{\cal D}={\cal D}{\sf D}={\cal D}$.
The numerator of the second pole term is given by
\begin{equation}\label{n32}
{\cal N}^{\mu\nu} ={\sf D}^{\mu\rho} p_\rho \frac{8  (  {p\!\!\!/}+2M)}{3p^2}  p_\sigma {\sf D}^{\sigma\nu}  \;.
\end{equation}
Both terms can easily be combined to produce
\begin{eqnarray}\label{mypropc}
G^{\mu\nu}  &=& {\sf D}^{\mu\rho} \bigg[ \frac{({p\!\!\!/}+M)g_{\rho\sigma}}{p^2-M^2}\nonumber\\
 & &{ }\quad\quad\quad\quad + \frac{2  p_\rho(  {p\!\!\!/}+2M)p_\sigma}{(p^2-M^2)(p^2-4M^2)}   \bigg] {\sf D}^{\sigma\nu} \;, 
\end{eqnarray}
which may be the preferred form for practical calculations.
This form clearly exhibits the fact that this propagator will only act in the space
of the RS fields $\psi^\mu_{\sf D}$ since 
\begin{equation}
\gamma_\mu G^{\mu\nu}=G^{\mu\nu}\gamma_\nu=0 \;.
\end{equation}
Moreover, it also shows that the limit of $p^2\rightarrow 0$ is well defined here since the
combination of the two pole terms cancels the individual $1/p^2$ singularities.

The construction given here leads to a clear separation of the spin-3/2 contributions proper and of
the additional spin-1/2 piece contained in the $(1,\frac{1}{2})$ RS fields. As is seen from Eq.\ (\ref{myprop}),
{\it the two contributions propagate with different masses!}

The first part of Eq.\ (\ref{myprop}),
\begin{equation}\label{wprop}
G^{\mu\nu}_{3/2}  = \frac{({p\!\!\!/}+M) {\cal D}^{\mu\nu}}{p^2-M^2}  \;,
\end{equation}
evidently contains only spin-3/2 in view of ${\cal D}^{\mu\nu}$ appearing here. This
corresponds to the propagator suggested by Williams \cite{williams},
which has been criticized \cite{benmer} because of the unphysical $1/p^2$ singularity 
appearing in ${\cal D}^{\mu\nu}$.
In view of the cancellations just described, the limit of vanishing $p^2$
poses no problem for Eq.\ (\ref{myprop}).

The second part,
\begin{equation}\label{nprop}
G^{\mu\nu}_{1/2}  = \frac{{\cal N}^{\mu\nu}}{p^2-(2M)^2}  \;,
\end{equation}
describes the propagation of the spin-1/2 part of the RS field $\psi^\mu_{\sf D}$, with a 
pole at $p^2=(2M)^2$. 
Furthermore, its numerator tensor contains an operator, $({p\!\!\!/}+2M)$,
which, for $p^2=(2M)^2$, can be expressed as 
%
\begin{equation}
{p\!\!\!/}+2M = 4M \sum U(p) \overline{U}(p)    \;, 
\end{equation}
i.e., it gives rise to an expansion in terms of Dirac spinors for the mass $2M$, where
\begin{equation}
({p\!\!\!/}-2M)U (p) = 0   \;. 
\end{equation}
Introducing the (unnormalized) function
%
\begin{eqnarray}\label{phifunc}
\phi^\mu(p) = {\sf D}^{\mu\nu}p_\nu U (p) &=& \left(  p^\mu-\sfrac{1}{4}\gamma^\mu {p\!\!\!/}  \right) U (p)\nonumber\\
            &=& \left(  p^\mu-\sfrac{M}{2}\gamma^\mu  \right) U (p)\; ,
\end{eqnarray}
suggested by the numerator of Eq.\ (\ref{n32}), one sees that
%
\begin{equation}
p_\mu \phi^\mu =  3 M^2 U(p)  \; ,
\end{equation}
apart from the unessential mass factor, is indeed a spin-1/2 spinor.
Moreover,
\begin{equation}
\left[{\sf D}_{\mu\nu}{p\!\!\!/}-Mg_{\mu\nu}\right]\phi^\nu(p)= 0 \; 
\end{equation}
plays the role of a wave equation.

The conclusion from this exercise is that in $(1,\frac{1}{2})$ spinor-vector space the
propagator $G^{\mu\nu}_{1/2}$ behaves very similar to the propagator
of a spin-1/2 particle. In any practical application, therefore, the pole of $G^{\mu\nu}_{1/2}$
at $p^2=(2M)^2$ may manifest itself in a manner which is largely indistinguishable 
from a spin-1/2 particle  of mass $2M$ in its own right.

As far as the general vertex (\ref{oldvertex}) is concerned, we immediately see that for the
present `singular' choice $A=-\sfrac{1}{2}$, the vertex becomes independent of $z$ and automatically
acquires a projector onto the $(1,\frac{1}{2})$ space, i.e.,
\begin{eqnarray}\label{newvertex}
\Gamma^\mu(q) \equiv \Gamma^\mu(q;-\sfrac{1}{2},z)  &=& \frac{f}{m_\pi} {\sf D}^{\mu\nu} q_\nu \nonumber\\
              &=& \frac{f}{m_\pi} \left( q^\mu -\sfrac{1}{4} \gamma^\mu {q\!\!\!/} \right) \;, 
\end{eqnarray}
which lends additional support to the construction of the propagator on the ${\sf D}$ 
subspace only given in Eq.\ (\ref{solveprop}). This vertex thus satisfies 
\begin{equation}\label{vertex_cond}
\gamma_\mu\Gamma^\mu(q) = 0 \; ,
\end{equation}
which is necessary in the present context.

Finally, if one wanted to eliminate $G^{\mu\nu}_{1/2}$ altogether, one would have to 
restrict the space even further and consider only the spin-3/2 subspace for which ${\cal D}^{\mu\nu}$
itself serves as the metric tensor. The resulting propagator then would be the Williams
propagator (\ref{wprop}), with its unphysical singularity at $p^2=0$ being just a reflection of the
fact that the metric of the space on which it is invertible is singular itself.
However, given the assumptions underlying the present approach, there is no basis for eliminating $G^{\mu\nu}_{1/2}$
in this manner. To be able to eliminate $G^{\mu\nu}_{1/2}$, it seems obvious that the condition (\ref{vertex_cond}) 
is not enough, but that one would need an interaction Lagrangian leading to the additional constraint
$p_\mu \Gamma^\mu=0$ which necessitates second-order derivatives in the interaction, 
as found already in Ref.\ \cite{pascalutsa}.

\subsection{Rarita--Schwinger case and \boldmath{$A=-\frac{1}{2}$}}

Since one can show quite generally that the $S$ matrix should not be affected by the values of $A$ in
(\ref{genlagrange}) \cite{kame}, we would like to add
some comments here regarding the effect of choosing $A=-\frac{1}{2}$ if one follows the procedure
described in Sec. \ref{sec:RS}.

As discussed already, for $A=-\frac{1}{2}$ the additional contact term $G_A^{\mu\nu}$ 
of Eq.\ (\ref{GRSA}) becomes infinite. However, the propagator never occurs by itself, but always appears between vertices. Considering therefore
\begin{equation}
\widetilde{G}_{A,\rho\sigma}=\theta_{\rho\mu}(A;z)G_A^{\mu\nu}\theta_{\nu\sigma}(A;z)  \;, 
\end{equation}
and writing
$A= -\frac{1}{2} +\varepsilon$, 
one finds a finite result,
\begin{eqnarray}
\widetilde{G}_A^{\mu\nu} &=& -\frac{1+4z}{12M^2}   \Big[\gamma^\mu p^\nu + p^\mu\gamma^\nu -\sfrac{1}{2}\gamma^\mu{p\!\!\!/}\gamma^\nu \nonumber\\
& &\quad{ }- \sfrac{1+4z}{8}\gamma^\mu ({p\!\!\!/}+2M)\gamma^\nu +\sfrac{1+4z}{2}M\gamma^\mu\gamma^\nu \Big]  \;, 
\end{eqnarray}
in the limit of $\varepsilon \to 0$. For $z=-\sfrac{1}{4}$, this term will vanish.
Therefore, since only
%
\begin{equation}
\theta^{\mu\nu}(-\sfrac{1}{2},-\sfrac{1}{4}) =\theta^{\mu\nu}(-1,-\sfrac{1}{4})={\sf D}^{\mu\nu} \; 
\end{equation}
will lead to identical matrix elements and hence an invariant description of observables,
this exercise clearly shows that one should choose $z=-\frac{1}{4}$. We mention that this is a value
compatible with the range of acceptable values found in fits to experimental data \cite{benmer}.

The value $z=-\sfrac{1}{4}$ had also been suggested in Ref.\ \cite{peccei} as being one for which all spin-1/2
contributions would vanish. The present derivation clearly shows that this is not the case; this choice
only restricts the allowed spin-1/2 contributions to those arising from the RS fields $\psi^\mu_{\sf D}$.

Comparing Eqs.\ (\ref{proprs}) and (\ref{myprop}),
as far as the final result is concerned, for $z=\frac{1}{4}$ both propagators
will act only in the subspace of the RS fields $\psi^\mu_{\sf D}$.
The reason they are not identical is due to the different routes of construction. While the propagator of
Eq.\ (\ref{myprop}) is constructed entirely within the space of the RS fields, the propagator
(\ref{proprs}) is derived on the entire space which contains both spin-1/2 contributions
and only the final result is projected onto the $(1,\frac{1}{2})$ subspace.
Recasting Eq.\ (\ref{mypropc}) in the form
\begin{equation}
G^{\mu\nu}  = {\sf D}^{\mu}_{\:\:\rho} \bigg[ \frac{({p\!\!\!/}+M)\Delta_{\text{RS}}^{\rho\sigma}}{p^2-M^2}
 + \frac{2  p^\rho(  {p\!\!\!/}+2M)p^\sigma}{3M^2(p^2-4M^2)}   \bigg] {\sf D}_\sigma^{\;\nu} \; 
\end{equation}
exhibits most clearly that this difference in the derivation leads to the appearance of the second pole term.

\section{Summary and Discussion}\label{sec:sum}

In summary, we have presented here a treatment of the free propagation of a massive spin-3/2 particle in which we use the 
so-called `singular' parameter values of the point transformations associated with the most general free Lagrangian 
containing first-order derivatives only. We have shown that in this way, we can easily eliminate the spin-1/2 contributions
which do not satisfy the Rarita--Schwinger constraint $\gamma_\mu \psi^\mu=0$. 

We should mention in this 
context that, with $\gamma_\mu \psi^\mu=0$ given, the other constraint, $p_\mu \psi^\mu=0$, follows from
the validity of the on-shell wave equation $({p\!\!\!/}-M)\psi^\mu=0$, of course.

Perhaps the most unexpected outcome of the present approach is the fact that a second pole term
{\it with a different mass} emerges from the procedure which isolates the spin-1/2 contribution.
In order to assess the ramifications of this finding for the description of physical observables, note
that, for $p^2=M^2$, the pole residues of $G^{\mu\nu}$, Eq.\ (\ref{myprop}), and $G^{\mu\nu}_{\text{RS}}$, 
Eq.\ (\ref{proprs}), are identical since $({p\!\!\!/}+M){\cal D}^{\mu\nu} \rightarrow ({p\!\!\!/}+M)\Delta^{\mu\nu}_{\text{RS}}$ in that
limit.  Taking the $\Delta(1232)$ as an example, this means that the well-established description of the
$\Delta$ properties based on the Rarita--Schwinger propagator is not going to change significantly
if described with the propagator of Eq.\ (\ref{myprop}) instead. 

Away from the first pole, in particular
near the second pole, this is no longer going to be true. It seems obvious that this second pole will
have a very similar effect on the numerical results as if one had added an additional, independent 
spin-1/2 particle with twice the original mass to the description of the problem of interest. 
Therefore, if the present results provide a valid description of physical reality, it means that
every observed spin-3/2 state that is due to an elementary field should be accompanied by 
a spin-1/2  state in the present approach. Taking the $P_{33}$ $\Delta(1232)$ as an example, one would have to look
for an $S_{31}$ or $P_{31}$ partner, depending on which of the spin-1/2 partial waves exhibits resonant behavior. 
The particle tables \cite{pdg98} provide a number of possible states.  Which, if any, of these resonances may be describable 
as the partner of the $\Delta(1232)$ according to the present formulation remains to be seen. 
The fact that their masses are  well below twice the $\Delta$ mass should not be
a problem, of course, since the present description applies to stable particles and does not take into
account dressing effects, different open decay channels, etc.  These effects can be expected to affect
the two poles in a
different way and therefore produce complex resonance pole positions no longer related in the simple manner
the bare poles are related on the real axis. A numerical study to answer these questions is underway. 

Whether the approach described here is realized in nature remains to be seen. However,
if the present description of spin-3/2 should turn out to be of practical relevance, it is
conceivable that a similar procedure may be applicable to even higher spin fields.

\acknowledgments
The author gratefully acknowledges useful discussions with Cornelius Bennhold and 
William C. Parke.
This work was supported in part by Grant No. DE-FG02-95ER-40907 of the 
U.S. Department of Energy.


\end{document}